\documentclass[final,3p,11pt]{elsarticle}

\usepackage{graphicx}%
\usepackage{dcolumn}%
\usepackage{bm}%
\usepackage {subeqnarray,epsfig,amsmath}
\usepackage{amssymb}

\makeatletter
\def\ps@pprintTitle{%
 \let\@oddhead\@empty
 \let\@evenhead\@empty
 \def\@oddfoot{}%
 \let\@evenfoot\@oddfoot}
\makeatother

\begin{document}

\begin{frontmatter}
\title{Extended Lubrication Theory: \\ Improved Estimates of Flow in Channels with Variable Geometry}

\author{Behrouz Tavakol}
\address{Wellman Center for Photomedicine, Mass. General Hospital, Harvard Medical School, Cambridge, MA 02139}
\address{Department of Biomedical Engineering \& Mechanics, Virginia Tech, Blacksburg, VA 24061}
\author{Guillaume Froehlicher}
\address{Department of Mechanical \& Aerospace Engineering, Princeton University, Princeton, NJ 08544}
\author{Douglas P. Holmes\corref{cor1}}
\ead{dpholmes@bu.edu}
\address{Department of Mechanical Engineering, Boston University, Boston, MA 02215}
\author{Howard A. Stone\corref{cor1}}
\ead{hastone@princeton.edu}
\address{Department of Mechanical \& Aerospace Engineering, Princeton University, Princeton, NJ 08544}
\cortext[cor1]{Corresponding authors}

\begin{abstract}
Lubrication theory is broadly applicable to the flow characterization of thin fluid films and the motion of particles near surfaces. We offer an extension to lubrication theory by starting with Stokes equations and considering higher-order terms in a systematic perturbation expansion to describe the fluid flow in a channel with features of a modest aspect ratio. Experimental results qualitatively confirm the higher-order analytical solutions while numerical results are in very good agreement with the higher-order analytical results. We show that the extended lubrication theory is a robust tool for an accurate estimate of pressure drop in channels with shape changes on the order of the channel height, accounting for both smooth and sharp changes in geometry.
\end{abstract}


\end{frontmatter}

\section{Introduction}
Lubrication theory is an approximation to the Navier-Stokes and continuity equations at low Reynolds numbers for narrow geometries with slow changes in curvature~\citep{Langlois1964,Leal2007, Ockendon}. The approach is used regularly to describe the velocity field and pressure gradient in fluid film lubricants~\citep{Szeri2005,Tichy2012}, the motion of particles within a fluid and near boundaries~\citep{GoldmanCoxBrenner,Stone2005}, the fluid flow passing through a microchannel with a known geometry~\citep{Amyot2007,Holmes2013,Plouraboue2004,Stone2004}, flow driven by the contracting walls of a soft channel~\citep{Tavakol2016}, $e.g.$, an insect's trachea~\citep{Aboelkassem2012,Aboelkassem2013}, and the flow of thin liquid films with  free surfaces~\citep{Oron,Snoeijer2006}, $e.g.$, when a droplet wets a solid surface~\citep{Bonn2009,Limat2004} or in spin coating over topographically patterned surfaces~\citep{Kalliadasis,StillwagonLarson}. 
Classical lubrication theory (CLT) is suitable for all of the above cases provided that the typical magnitude of the boundary slope is sufficiently small, typically on the order of $\mathcal{O}\left (10^{-1}\right )$ or less. Due to its simplicity and versatile applications, lubrication theory is widely applied and it is appreciated that it often works beyond its formal limits of validity. The ideas have even been extended to consider inertial effects, e.g. \citet{Wilson1998}. Also, earlier work on flows in sinusoidally constricted pipe with radial variations comparable to axial variations was studied numerically, e.g. \citet{Tilton1984}, where we note that our higher-order approach below effectively gives an analytical solution to the problem. 

In this study, we obtain higher-order terms of the lubrication approximation and present an extension to lubrication theory, which we refer to as extended lubrication theory (ELT), to address two limitations of CLT. First, the use of ELT is no longer limited to small gaps and thin films. Second, the boundaries can be described by any mathematical shape function with modest curvatures as long as they are continuous and differentiable. 
In addition, we show how the differentiability condition may be relaxed at low Reynolds numbers, at least in practice, by considering geometries that are piece-wise differentiable. We compare the results of different orders of the analytical solutions with experimental results and also with direct numerical solutions of the Navier-Stokes equations to define a threshold for considering higher-order terms in the solution. The theory will breakdown for sufficiently large wall slopes and then if an analysis is desired, an approach with matched asymptotic expansions is necessary, e.g. \citet{CormackLeal}. Note that a related idea emphasizing the first correction to the classical lubrication approach, but focused on the mathematical form of the differential equations, is given by \citet{MarusicPaloka}.

Our approach here can be used to describe the flow in a channel with a large constriction made by a trapped cell, particle, bubble, or droplet, or as a result of buckled, crumpled, or swollen walls. We validate this approach for large non-uniformities within the fluid channel by varying two characteristic geometric parameters, the constriction's length and amplitude, from zero to the channel height, e.g. the corresponding dimensionless geometric parameters vary from zero to one. We note that in addition to numerical solutions, two types of analytical calculations have been considered in the literature. One method is domain perturbation \citep{HASEGAWA1983,Hemmat1995,Malevich2006,Manton1971,Pozrikidis1987} where the constriction amplitude is much smaller than the channel height. The second method involves corrections to leading-order lubrication theory \citep{Disimile1983,Plouraboue2004,Wang1978,Wang2004} while the constriction amplitude can be as high as the channel height. We show the applicability of our approach to different geometries and provide the accuracy of such results as two nondimensional geometric parameters along with Reynolds number vary.  Finally, we note that a similar higher-order expansion was used recently in a study of an electrokinetic flow in a channel of nonuniform shape \citep{Michelin2015}.

\section{Theoretical Approximation}

\begin{figure}[t]
\begin{center}
\resizebox{0.5\columnwidth}{!} {\includegraphics{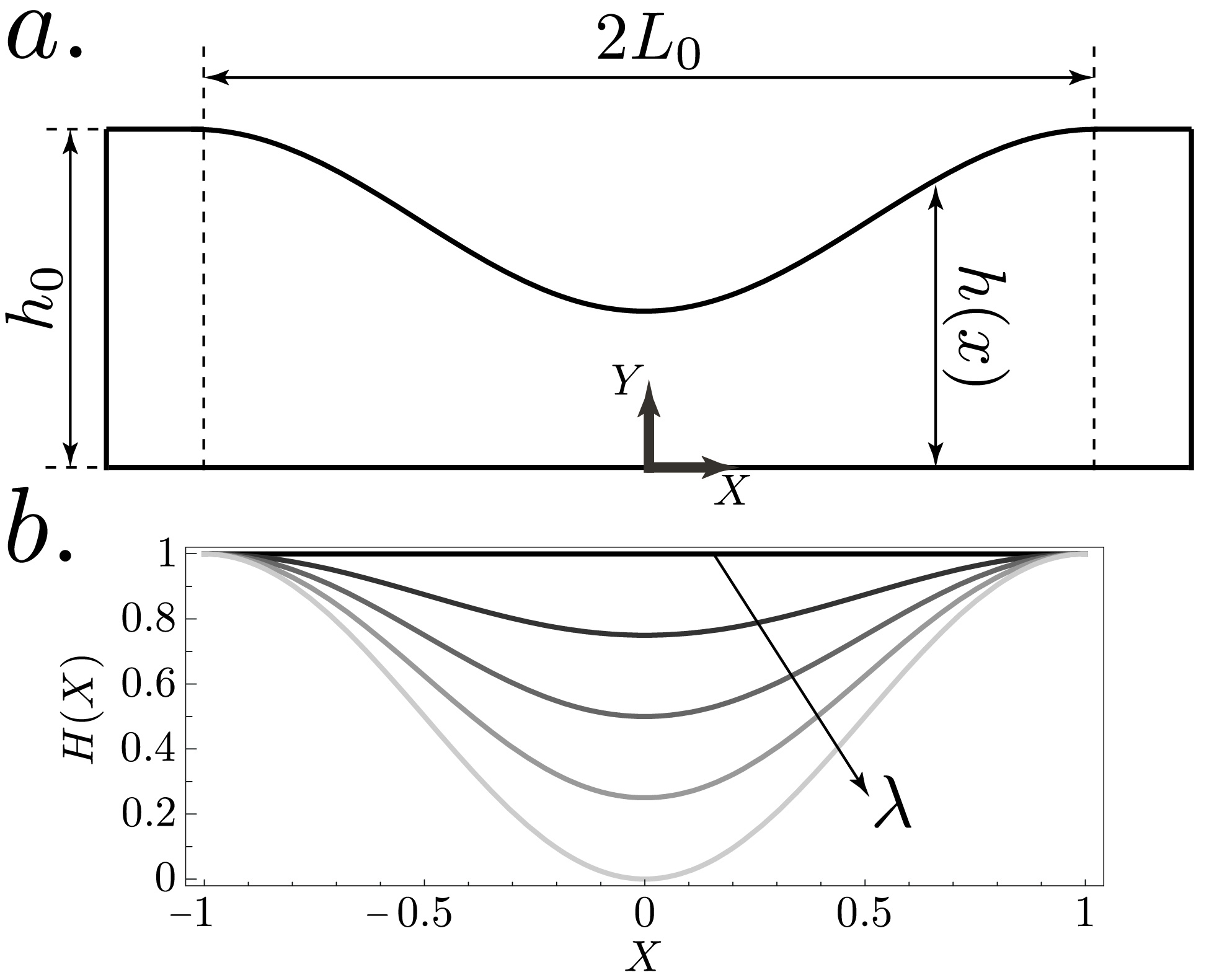}}
\end{center}
\vspace{-5mm}
\caption[]{{\sl (a)} A schematic of the channel with shape $y=h(x) = h_0 H(x)$. {\sl (b)} Shape function of $  H(X) = 1 - \frac{\lambda}{2} \left(1+\cos\left (\pi X\right )\right)$ for different $\lambda$.
\label{fig:shapefunction}}
\vspace{-2mm}
\end{figure}

We consider working to higher order in traditional lubrication theory to describe fluid flow in nonuniform channel shapes with modest aspect ratios. Thus, we consider incompressible, steady, two-dimensional pressure-driven flow in a channel with shape $y=h(x)=h_0 H(X)$, where $X=x/L_0$, $L_0$ is the channel length, $h_0$ is a characteristic channel height, $H(X)$ is a normalized shape function, and $\delta = h_0/L_0\ll 1$. A typical geometry in the form of a constriction is shown in Fig.~\ref{fig:shapefunction}a. We assume that the Reynolds number is small and so consider the continuity and Stokes equations
\begin{equation}
\nabla\cdot{\bf u} = 0 \qquad\hbox{and}\qquad \mu\nabla^2 {\bf u} = \nabla p,
\label{eq:stokes}
\end{equation}
where ${\bf u} = (u,v)$ is the velocity field and $\mu$ is the fluid viscosity. We denote the constant flow rate (per unit width) as $q_0$. A discussion of the flow based on the full Navier-Stokes equations is given in \S \ref{NumericalSolutions}.

Consistent with the traditional lubrication approximation we analyze equation (\ref{eq:stokes}) and choose to introduce dimensionless variables according to
\begin{equation}
X=\frac{x}{L_0}, \quad Y=\frac{y}{h_0}, \quad U=\frac{u}{q_0/h_0}, \quad V=\frac{v}{q_0/L_0},\quad P=\frac{p}{\Delta p}=\frac{p}{\mu q_0 L_0/h_0^3 }.
\label{eq:nondimparam}
\end{equation}
Note that the scaling for the transverse velocity component $v$ is $\mathcal{O}(h_0/L_0)\ll 1$ smaller than the scaling for $u$, which is the component of flow along the channel axis.
The dimensionless equations corresponding to~\eqref{eq:stokes} for a two-dimensional flow are

\begin{subequations}
\begin{align}
\frac{\partial U}{\partial X}+\frac{\partial V}{\partial Y}&=0\\
\delta^2 \frac{\partial^2 U}{\partial X^2}+\frac{\partial^2 U}{\partial Y^2}&=\frac{\partial P}{\partial X}\\
\delta^4 \frac{\partial^2 V}{\partial X^2}+\delta^2\frac{\partial^2 V}{\partial Y^2}&=\frac{\partial P}{\partial Y}.
\end{align}
\label{eq:nondimstokes}
\end{subequations}These equations are to be solved with boundary conditions
\vspace{-3mm}
\begin{equation}
U=0, \,\,V=0 \quad\hbox{at $Y=0, H(X)$}, \quad\hbox{and}\quad \int_0^{H(X)} U(X,Y)~{\rm d}Y = 1,
\vspace{-2mm}
\label{eq:bcs}
\end{equation}
where the integral constraint states that the total flow rate is prescribed. We will determine the corresponding pressure drop across the constriction such that the pressure gradient tends to a constant as $X\rightarrow \pm 1$. Note that the problem statement only involves one dimensionless parameter $\delta^2$. However, we later introduce another dimensionless parameter, $\lambda$, coming from the shape function and corresponding to the normalized constriction amplitude. Also, we will show below that for sinusoidal shape variations with dimensionless amplitude $\lambda$ (see Fig. \ref{fig:shapefunction}b) that the effective small parameter is $\left (\delta \lambda\right )^2$, corresponding to the square of the typical wall slope.

\subsection{Perturbation expansion and the leading-order result from lubrication theory}

Our first steps follow standard discussions in textbooks, e.g. Leal \citep{Leal2007}. 
Because the problem statement only involves  $\delta^2$, which is assumed to be small, we seek a solution to~\eqref{eq:nondimstokes} of the form
\begin{subequations}
\begin{align}
U(X,Y; \delta) &= U_0(X,Y) + \delta^2 U_2 (X,Y) + \delta^4 U_4 (X,Y)+\cdots\\
V(X,Y; \delta) &= V_0(X,Y) + \delta^2 V_2 (X,Y) + \delta^4 V_4 (X,Y)+\cdots\\
P(X,Y; \delta) &= P_0(X,Y) + \delta^2 P_2 (X,Y) + \delta^4 P_4 (X,Y)+\cdots
\end{align}
\end{subequations}

At leading order, we have the familiar classical lubrication problem
\begin{subequations}
\begin{align}
\frac{\partial U_0}{\partial X}+\frac{\partial V_0}{\partial Y}&=0 \label{continuityleadingorder}\\
\frac{\partial^2 U_0}{\partial Y^2}&=\frac{\partial P_0}{\partial X}\\
\frac{\partial P_0}{\partial Y}&=0, 
\end{align}
\end{subequations}with $U_0=0$ at $Y=0 \text{ and } H(X)$. The solution is
\begin{equation}
U_0(X,Y) = \frac{1}{2}\frac{d P_0}{d X}\left ( Y^2-YH(X)\right )
\end{equation}and the pressure gradient, which only depends on $X$, follows from applying the integral constraint $\left (\int_0^{H(X)} U_0(X,Y)~{\rm d}Y=1\right )$

\begin{equation}
\frac{d P_0}{d X}=-\frac{12}{H(X)^3}.
\end{equation}The corresponding velocity distribution is then
\begin{equation}
U_0(X,Y)= \frac{6}{H(X)^3}\left( YH(X)-Y^2\right ).
\label{ExtendedLubricationVelocityLeadingOrder}
\end{equation}

To provide an example, we consider the shape function
\begin{equation}
H(X)= 1 - \frac{\lambda}{2} \left ( 1+\cos\left (\pi X\right )\right )\quad\left (0\le \lambda < 1\right ),
\label{eq:ShapeFunction}
\end{equation} 
as sketched in Fig.~\ref{fig:shapefunction}b. Here $\lambda$ corresponds to the dimensionless amplitude of the change in the channel height. Thus, the wall slope has magnitude $\left | \frac{dh}{dx}\right | = O\left ( \frac{h_0}{L_0}\lambda\right ) = O(\delta \lambda)$. 

The leading-order pressure drop $\Delta P_0$ is then calculated to be  (the integration was accomplished with Mathematica)
\begin{align}
P_0(-1)-P_0(1)=\Delta P_0 & = -\int_{-1}^1 \frac{d P_0}{d X}
~{\rm d}X =12\int_{-1}^1
\frac{1}{\left(1 - \frac{\lambda}{2} -\frac{\lambda}{2}\cos\left (\pi X\right )\right)^3}~{\rm d}X \nonumber \\ 
&= \frac{3\left (3\lambda^2-8\lambda+8\right )}{\left (1-\lambda\right )^{5/2}}.
\label{Eq:DelP0}
\end{align}The pressure drop becomes singular in the limit $\lambda \rightarrow 1$ as the gap vanishes.

Before proceeding further, we determine the velocity component $V_0 (X,Y)$ using the continuity equation. Although equation \eqref{continuityleadingorder} is first order in $Y$, we expect it to satisfy two boundary conditions, as $V_0 (X,0)=V_0(X,H(X))=0$. 
Using the continuity equation, and imposing $V_0\left(X,0\right)=0$, we have 
\begin{equation}
V_0(X,Y) = -\int_0^{Y} \frac{\partial U_0\left(X,S\right)}{\partial X}~{\rm d}S,
\end{equation} which yields
\begin{equation}
V_0(X,Y) = 2Y^3 \left (H^{-3}\right )^\prime - 3Y^2 \left (H^{-2}\right )^\prime,
\label{VvelocityLeadingOrder}
\end{equation}where primes denotes $X$ derivatives. We then note that at $Y=H(X)$ direct differentiation shows that (\ref{VvelocityLeadingOrder}) yields $V_0(X,H(X))=0$. Alternatively, we can write
\begin{equation}
V_0(X,H(X)) = -\int_0^{H(X)} \frac{\partial U_0}{\partial X}~{\rm d}Y=-\frac{d}{dX}
\int_0^{H(X)}  U_0~{\rm d}Y+U_0(X,H(X)) \frac{dH}{dX}=0,
\end{equation}as the second term on the right-hand side vanishes owing to the no-slip condition and the first term on the right-hand side vanishes since the flow rate is constant. The same idea applies for evaluating the $Y$-component of velocity at every order in the analysis below and the no-slip boundary condition is satisfied for both velocity components of $\bf u$. 

\subsection{The $\mathcal{O}(\delta^2)$ term in the perturbation expansion}

In most calculations utilizing lubrication theory the development is truncated with the leading-order term calculated in the preceding section, which is well-known from countless applications. Here our interest is to improve the approximation by including additional terms in the perturbation solution. 
At the next order, $\mathcal{O}\left (\delta^2\right )$, the perturbation expansion yields
\begin{subequations}
\begin{align}
\frac{\partial U_2}{\partial X}+\frac{\partial V_2}{\partial Y}&=0\\
\frac{\partial^2 U_2}{\partial Y^2}-\frac{\partial P_2}{\partial X}&=-\frac{\partial^2 U_0}{\partial X^2}\\
\frac{\partial P_2}{\partial Y}&=\frac{\partial^2 V_0}{\partial Y^2},
\end{align}
\label{ExtendedLubricationSecondOrder}
\end{subequations}
\hspace{-0.5em}with boundary conditions $U_2=0$ at $Y=0 \text{ and } H(X)$, and $\int_0^{H(X)} U_2 (X,Y)~{\rm d}Y=0$.
This last integral constraint follows since all of the fluid flux is specified in the scaling used to establish the leading-order problem. We seek the velocity distribution and pressure drop $\Delta P_2 = P_2\left(-1\right)-P_2\left(1\right)$ needed to enforce the constraint on the flux.

We can integrate the last equation of \eqref{ExtendedLubricationSecondOrder}, and use continuity, to obtain
\begin{equation}
P_2(X,Y) = -\frac{\partial U_0}{\partial X} + c_3 (X),
\label{ExtendedLubricationPressureEquation2}
\end{equation}where the function $c_3(X)$ is allowed by the integration. With this pressure distribution, we use the $X$-momentum equation to find
\begin{equation}
\frac{\partial^2 U_2}{\partial Y^2}=\frac{d c_3}{d X}
-2\frac{\partial^2 U_0}{\partial X^2}
\end{equation}where $U_0$ is given by equation (\ref{ExtendedLubricationVelocityLeadingOrder}). Upon integration, and application of the boundary conditions, we find
{\small
\begin{align}
U_2(X,Y)=&-2\left (H(X)^{-2}\right )^{\prime\prime}\left (Y^3 -H(X)^2Y\right )\nonumber \\
&+\left (H(X)^{-3}\right )^{\prime\prime}\left (Y^4 -H(X)^3Y\right )
+\frac{1}{2}\frac{d c_3}{d X}\left ( Y^2 -H(X)Y\right ),
\label{U2equation1}
\end{align}}where primes indicate derivatives with respect to $X$.

Since we have accounted for the specified dimensionless flow rate at
leading order, then we now require  $\int_0^{H(X)} U_2 (X,Y)~{\rm d}Y=0$, which leads to
\begin{equation}
\frac{d c_3}{dX} = 6\left (H(X)^{-2}\right )^{\prime\prime}H(X)-
\frac{18}{5}\left (H(X)^{-3}\right )^{\prime\prime}H(X)^2.
\label{ExtendedLubricationHigherOrderPressure2}
\end{equation}
Equations \eqref{U2equation1} and \eqref{ExtendedLubricationHigherOrderPressure2} give the second-order $X$-component of the velocity $U_2 (X,Y)$ for any shape function $H(X)$. To continue with the example of Fig. \ref{fig:shapefunction}, we again use the shape function in equation \eqref{eq:ShapeFunction}. Integrating \eqref{ExtendedLubricationPressureEquation2}, 
 taking into account that $\frac{\partial U_0}{\partial X}$ vanishes as $X\rightarrow -1 \text{ and }1$,
 and using \eqref{ExtendedLubricationHigherOrderPressure2}, we obtain the pressure drop $\Delta P_2$ at this order as
\begin{equation}
P_2(-1)-P_2(1)=\Delta P_2 = -\int_{-1}^1 \frac{\partial P_2}{\partial X}~{\rm d}X =
-\int_{-1}^1 \frac{d c_3}{d X}~{\rm d}X= \frac{12\pi^2\lambda^2}{5\left (1-\lambda\right )^{3/2}}.
\label{Eq:DelP2}
\end{equation}

We determine $V_2(X,Y)$ using the continuity equation and imposing $V_2(X,0)=0$, which leads to the expression
\begin{align}
V_2(X,Y)&=\left(H(X)^{-2}\right)^{\prime\prime\prime} \left ( \frac{1}{2}Y^4-H(X)^2 Y^2\right ) -2\left(H(X)^{-2}\right)^{\prime\prime}H^{\prime}(X) Y^2\nonumber\\
&-\left(H(X)^{-3}\right)^{\prime\prime\prime} \left ( \frac{1}{5}Y^5-H(X)^3 Y^2\right ) + \frac{3}{2} \left(H(X)^{-3}\right)^{\prime\prime}\left(H^{\prime}(X)\right)^2 Y^2\nonumber\\
&-\frac{d^2c_3}{dX^2} \left (\frac{1}{6}Y^3-\frac{1}{4}H Y^2\right )+\frac{1}{4}\frac{dc_3}{dX}H^{\prime}(X)Y^2.
\label{Eq:V2}
\end{align}

This equation only involves the shape function $H(X)$, since $\frac{dc_3}{dX}$ is given in (\ref{ExtendedLubricationHigherOrderPressure2}). As in the previous section, it can be verified that $V_2(X,H(X))=0$.

We conclude this subsection by noting that we have now established the net pressure drop at this order:
\begin{align}
\Delta P &= \Delta P_0 +\delta^2 \Delta P_2 + O\left (\delta^4\right ) \nonumber\\
&=\frac{3\left (3\lambda^2-8\lambda+8\right )}{\left (1-\lambda\right )^{5/2}}+ \frac{12\left (\pi \delta \lambda\right )^2}{5\left (1-\lambda\right )^{3/2}} + O\left (\delta^4\right ).
\end{align}
This result emphasizes that for small slopes of the boundaries, then the correction to classical lubrication theory is $O
\left (\left (\delta \lambda\right )^2\right )$, which corresponds to the square of the wall slope.

\subsection{The perturbation expansion at $\mathcal{O}\left (\delta^4\right)$}

It is useful to go one step further simply to illustrate that the basic analytical steps carry through at every order. The higher-order terms help to provide a better representation of flows in geometries with more rapid shape variations.
We can continue these basic steps at $\mathcal{O}\left (\delta^4\right )$, where we have the equations
\begin{subequations}
\begin{align}
\frac{\partial U_4}{\partial X}+\frac{\partial V_4}{\partial Y}&=0\\
\frac{\partial^2 U_4}{\partial Y^2}-\frac{\partial P_4}{\partial X}&=- \frac{\partial^2 U_2}{\partial X^2}\\
\frac{\partial P_4}{\partial Y}&=\frac{\partial^2 V_2}{\partial Y^2}+\frac{\partial^2 V_0}{\partial X^2},
\end{align}
\end{subequations}
with $U_4=0$ at $Y=0 \text{ and }H(X)$, and $\int_0^{H(X)} U_4 (X,Y)~{\rm d}Y=0$.
Using the results obtained above, these equations can be solved, though the algebraic manipulations involved become progressively more cumbersome. We outline the main steps below. First, the $Y$-momentum equation can be integrated, which, after using the continuity equation, yields
\begin{equation}
P_4 (X,Y)= -\frac{\partial U_2}{\partial X} + \frac{\partial^2}{\partial X^2}
\int_0^Y V_0 (X,S)~{\rm d}S + c_5 \left(X\right).
\end{equation}Second, from the $X$-momentum equation we have
\begin{equation}
\frac{\partial^2 U_4}{\partial Y^2}= -2\frac{\partial^2 U_2}{\partial X^2}
+\frac{\partial^3}{\partial X^3}
\int_0^Y V_0 (X,S)~{\rm d}S + \frac{dc_5}{dX}.
\end{equation}Since $U_2(X,Y)$ is known from equation (\ref{U2equation1}), then we calculate
\begin{align}
\frac{\partial^2 U_2}{\partial X^2} &= \left ( H^{-3}\right )^{\prime\prime\prime\prime} Y^4
-2 \left ( H^{-2}\right )^{\prime\prime\prime\prime} Y^3
+\left [ -\left (H^3\left ( H^{-3}\right )^{\prime\prime}\right )^{\prime\prime} 
+2 \left (H^2\left ( H^{-2}\right )^{\prime\prime}\right )^{\prime\prime} \right ] Y \nonumber \\
&+ \frac{1}{2}\frac{d^3c_3}{dX^3} Y^2 -\frac{1}{2}\left (H\frac{dc_3}{dX}\right )^{\prime\prime} Y.
\end{align}
Combining the last two results, we find
\begin{align}
\frac{\partial^2 U_4}{\partial Y^2}&=  -\frac{3}{2}
\left ( H^{-3}\right )^{\prime\prime\prime\prime} Y^4
+3 \left ( H^{-2}\right )^{\prime\prime\prime\prime} Y^3
+\left [ 2\left (H^3\left ( H^{-3}\right )^{\prime\prime}\right )^{\prime\prime}
-4 \left (H^2\left ( H^{-2}\right )^{\prime\prime}\right )^{\prime\prime} \right ] Y \nonumber \\
&- \frac{d^3c_3}{dX^3} Y^2 +\left (H\frac{dc_3}{dX}\right )^{\prime\prime} Y + \frac{dc_5}{dX}.
\end{align}
It is straightforward to integrate twice and apply $U_4=0$ at $Y=0\text{ and }H(X)$ to arrive at 
\begin{align}
U_4 (X,Y) &= -\frac{1}{20}\left ( H^{-3}\right )^{\prime\prime\prime\prime}\left (Y^6-H^5 Y\right ) +\frac{3}{20} \left ( H^{-2}\right )^{\prime\prime\prime\prime}
\left (Y^5-H^4 Y\right ) \nonumber \\
&+ \frac{1}{3}\left [ \left (H^3\left ( H^{-3}\right )^{\prime\prime}\right )^{\prime\prime}
-2 \left (H^2\left ( H^{-2}\right )^{\prime\prime}\right )^{\prime\prime} \right ]
\left (Y^3-H^2 Y\right ) \nonumber \\
&- \frac{1}{12}\frac{d^3c_3}{dX^3}\left (Y^4-H^3 Y\right )  +
\frac{1}{6}\left (H\frac{dc_3}{dX}\right )^{\prime\prime} \left (Y^3-H^2 Y\right ) + \frac{1}{2}\frac{dc_5}{dX}\left (Y^2-H Y\right ).
\label{eq:U4}
\end{align}
Since $\int_0^{H(X)} U_4 (X,Y)~{\rm d}Y=0$, we obtain $\frac{dc_5}{dX}$:
\begin{align}
\frac{dc_5}{dX}&=  \frac{3}{14}
\left ( H^{-3}\right )^{\prime\prime\prime\prime} H^4
-\frac{3}{5} \left ( H^{-2}\right )^{\prime\prime\prime\prime} H^3
-\left [ \left (H^3\left ( H^{-3}\right )^{\prime\prime}\right )^{\prime\prime}
-2 \left (H^2\left ( H^{-2}\right )^{\prime\prime}\right )^{\prime\prime} \right ] H \nonumber \\
&+ \frac{3}{10}\frac{d^3c_3}{dX^3} H^2 -\frac{1}{2}\left (H\frac{dc_3}{dX}\right )^{\prime\prime} H.
\label{eq:dc5dx}
\end{align}

The equations \eqref{ExtendedLubricationHigherOrderPressure2}, \eqref{eq:U4}, and \eqref{eq:dc5dx} give the $X-$component velocity at this order for any choice of the shape function $H(X)$. We determine the correction to the pressure drop as 
\begin{align}
P_4(-1)-P_4(1)=\Delta P_4 &= -\int_{-1}^1 \frac{\partial P_4}{\partial X}~{\rm d}X =
-\int_{-1}^1 \frac{d c_5}{d X}~{\rm d}X \nonumber \\ &= \-\frac{8\pi^4\left (-428\left(-1+\sqrt{1-\lambda}\right)+214\left(-2+\sqrt{1-\lambda}\right)\lambda+53\lambda^2\right )}{175\sqrt{1-\lambda}},
\label{Eq:DelP4}
\end{align}
where we have used Mathematica to accomplish the final integration for the shape function \eqref{eq:ShapeFunction}.

For a given flow rate $(q_0)$, we have determined the dimensionless pressure drop $\Delta P = $ \\$\left(\Delta p_{\,measured}\right)/\left(\mu \,q_0 L_0/h_0^3\right)$ as a function of the shape parameters $\delta$ and $\lambda$, where $\Delta p_{\,measured}$ is the difference in pressure measured at the two ends of the constriction. In particular, $\Delta P = \Delta P_0(\lambda)+\delta^2 \Delta P_2(\lambda) +\delta^4 \Delta P_4(\lambda) + \mathcal{O}(\delta^6)$, where $\lambda$ is defined by the given shape function (\ref{eq:ShapeFunction}).

We next describe experiments and numerical simulations to confirm the improved description offered by these additional terms in the lubrication approximation.

\begin{figure}
\begin{center}
\resizebox{1\columnwidth}{!} {\includegraphics{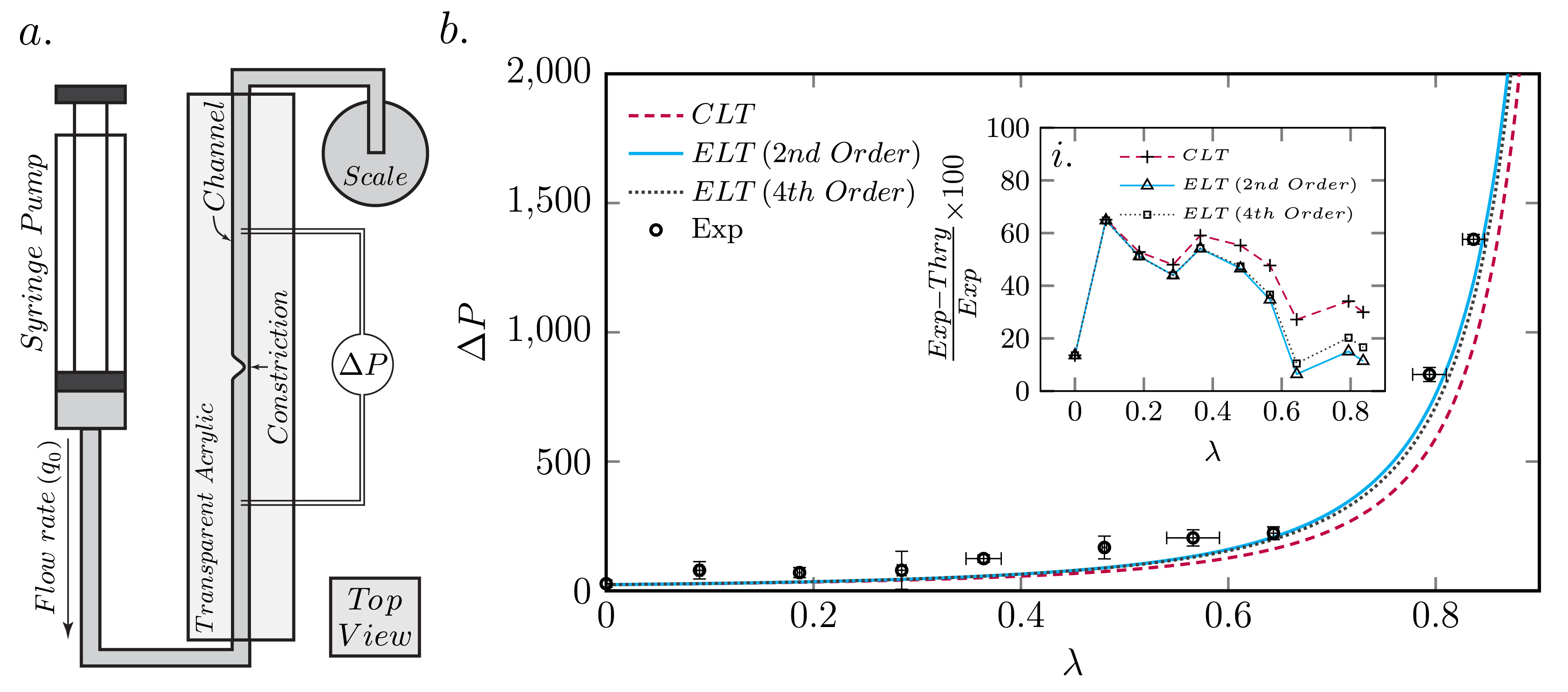}}
\end{center}
\vspace{-5mm}
\caption[]{{\sl a.} Schematic of the experimental setup from the top view. A constant flow rate was applied using a syringe pump and the pressure drop was measured using a sensitive pressure sensor. A scale was also used to verify the applied flow rate. For all of the experiments, $L_0\approx h_0 \approx5\,\text{mm}$. {\sl b.} Comparison of the dimensionless pressure drop across the arch showing the analytical solutions and experimental results ($\delta \approx 1$). The inset shows the difference between the experiment and the theory, suggesting that higher orders of the analytical solutions are in better agreement with the experimental results. \vspace{-2mm}
\label{fig:Exp}}
\end{figure}

\section{Experimental Verification}

Our experimental setup consists of a long channel (200 mm) with a rectangular cross section (5 mm by 5 mm) and an obstruction in the middle (Fig. \ref{fig:Exp}a), with $L_0\approx h_0$. The sides of the channel were cut from acrylic sheets (8560K211, McMaster-Carr) using a laser cutter (Epilog Mini Laser 24, 60 Watts) and bonded together using an acrylic capillary cement (10705, TAP Plastics). We varied the arch size and shape, similar to Fig. \ref{fig:shapefunction}, while keeping all other geometrical parameters constant between different tests. We recognize that our theory is two dimensional and the experimental geometry is three dimensional. However, as the arch amplitude increases, the flow through the narrow gap better approximates a two-dimensional flow, which allows us to highlight the impact on the pressure drop of systematically increasing the magnitude of the shape variations.

The pressure drop within the channel between two fixed points, symmetrically located on each side of the arch, was measured using a sensitive differential pressure sensor (CPCL04D, Honeywell). By keeping the flow rate constant for different tests, the pressure drop across the arch was then obtained by subtracting the pressure drop within the flat part of the channel from the total pressure drop between those fixed points. The fluid was chosen to be a standard viscosity oil (N1000, Cannon Instrument) and the temperature was kept at $22.5\pm0.5 ^{o}\text{C}$, resulting in a viscosity of 2.45 $\text{Pa}.\text{s}$ and density of 848 $\text{kg}/\text{m}^3$. We used a syringe pump (PHD Ultra CP, Harvard Apparatus) to apply a fixed flow at a rate of 14.4 $\text{mm}^3/\text{s}$ so that $\mathcal{R}e \approx 0.001$ for all the tests. The arch height was measured by taking images of the setup using a camera (FASTCAM Mini UX100) and then processing those images using a MATLAB code. 

We repeated each test at least three times and the average measured pressure drops along with their associated standard deviations are shown in Fig. \ref{fig:Exp}b, where $\delta \approx 1$ and $\lambda$ is varied from zero (flow in a straight channel) to $\lambda\approx 0.9$. We note that higher orders of the analytical solutions are in better agreement with the experimental results while the CLT (red dotted line) underestimates the results by about 40 \%. We also note that the analytical solutions are obtained in a two-dimensional channel while the experimental results are from a three-dimensional channel, and this is one of the reasons for the difference between the analytical and experimental results. 
Although the experimental results confirm the trend in the analytical solutions, it may not be feasible to conduct experiments for every case to verify the higher orders of the analytical solutions due to the sensitivity and the difficulty of such experiments. So we performed numerical simulations for variety of cases and compared those results with the analytical solutions in the next section.  

\section{Numerical Simulations}
\label{NumericalSolutions}

As an additional route to test the analytical results above, 
we seek to numerically solve the Navier-Stokes equation in their full form for incompressible, steady two-dimensional flow. Again, we assume a specified flow rate and seek the corresponding pressure drop as geometric parameters are varied. The same scalings and dimensionless parameters introduced in equation \eqref{eq:nondimparam} are used to obtain the dimensionless continuity and Navier-Stokes equations
\begin{subequations}
\begin{align}
\frac{\partial U}{\partial X}+\frac{\partial V}{\partial Y}&=0\\
\mathcal{R}e\,\, \delta \left(U \frac{\partial U}{\partial X}+V\frac{\partial U}{\partial Y}\right) &=-\frac{\partial P}{\partial X}+\delta^2 \frac{\partial^2 U}{\partial X^2}+\frac{\partial^2 U}{\partial Y^2}\\
\mathcal{R}e\,\, \delta^3 \left(U \frac{\partial V}{\partial X}+V\frac{\partial V}{\partial Y}\right)& =-\frac{\partial P}{\partial Y}+\delta^4 \frac{\partial^2 V}{\partial X^2}+\delta^2\frac{\partial^2 V}{\partial Y^2},
\end{align}
\label{eq:nondimNSsim}
\end{subequations}
\hspace{-0.5em}where $\mathcal{R}e = \rho \,q_0/ \mu$ is the Reynolds number, $\rho$ is the fluid density, and $\mu$ is the fluid viscosity. In the lubrication literature, it is common to define the reduced Reynolds number, ${\mathcal{R}e}^\ast= \mathcal{R}e \,\delta$, which also appears in \eqref{eq:nondimNSsim}. Equation (\ref{eq:nondimNSsim}) emphasizes that there are two parameters needed to describe the flow, $\delta$ and $\mathcal{R}e$, and a third parameter, $\lambda$, enters through a specification of the channel shape.

A numerical solver often uses the weak form of \eqref{eq:nondimNSsim}. To do so, we consider an arbitrary pair of $P$ and ${\bf U} = (U,V)$ to be a solution to the dimensionless continuity and Navier-Stokes equations (\ref{eq:nondimNSsim}) for a steady and incompressible flow. If these equations are multiplied by any pressure and velocity basis functions, $i.e.$ $(q, \nu_1, \nu_2)$, and integrated over the domain $\Omega$, the pair is still a solution, and satisfies the new equations. We then reduce all the second-order terms to first-order ones using Gauss's theorem and neglect the boundary integrals as they are usually handled separately in finite element packages. Therefore, we have the following weak form of the continuity and Navier-Stokes equations that can directly be used within numerical solvers:
\begin{subequations}
\begin{align}
0 = & \int_{\Omega}{\left[\frac{\partial U}{\partial X}+\frac{\partial V}{\partial Y}\right] q~{\rm d}\Omega}\\
0 = & \int_{\Omega}{\left[\nu_1\left(\mathcal{R}e \, \delta \left(U \frac{\partial U}{\partial X}+V \frac{\partial U}{\partial Y} \right) + \frac{\partial P}{\partial X} \right)+\delta^2 \frac{\partial \nu_1}{\partial X} \frac{\partial U}{\partial X}+ \frac{\partial \nu_1}{\partial Y} \frac{\partial U}{\partial Y}\right]~{\rm d}\Omega} \\
0 = & \int_{\Omega}{\left[\nu_2\left(\mathcal{R}e \, \delta^3 \left(U \frac{\partial V}{\partial X}+V \frac{\partial V}{\partial Y} \right)+ \frac{\partial P}{\partial Y} \right)+ \delta^4 \frac{\partial \nu_2}{\partial X} \frac{\partial V}{\partial X}+ \delta^2 \frac{\partial \nu_2}{\partial Y} \frac{\partial V}{\partial Y}\right]~{\rm d}\Omega}.
\end{align}
\label{eq:weakform_NS}
\end{subequations}

\begin{figure}
\begin{center}
\resizebox{0.75\columnwidth}{!} {\includegraphics{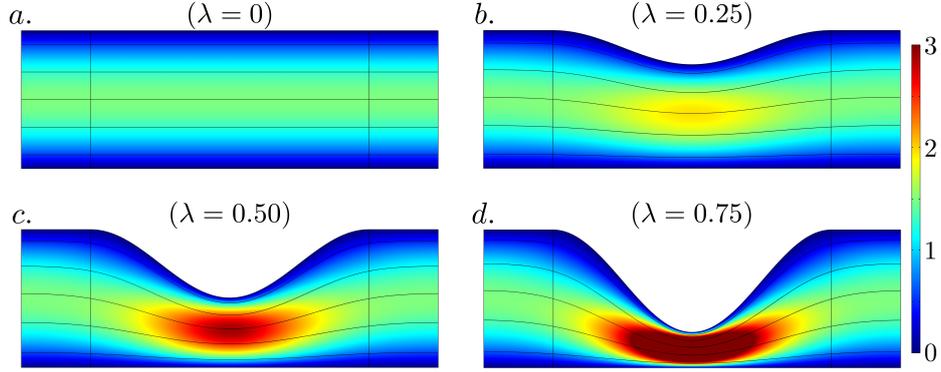}}
\end{center}
\vspace{-5mm}
\caption[]{Velocity magnitude obtained using numerical simulations for $\mathcal{R}e=1, \delta=1$, and $\lambda$ varied from 0 to 0.75 ({\sl a}-{\sl d}). 
\label{fig:SimResults}}
\vspace{-2mm}
\end{figure}

We used COMSOL 4.3b mainly as a PDE solver to the above equations. These dimensionless equations depend on two dimensionless parameters: Reynolds number and the geometric variable $\delta = h_0/L_0$. In addition, the shape function \eqref{eq:ShapeFunction} adds another dimensionless parameter, $\lambda$, to this system. 
So we performed the numerical simulations first for $\mathcal{R}e=0$ while varying $\lambda$ from $0$ to $0.99$ and $\delta$ from $0.2$ to $1$. 
The dimensionless channel height was $H_0=1$ and we applied a flow at the inlet at a fixed rate with a parabolic velocity profile of $6(Y-Y^2)$ so that $Q=1$. The outlet pressure was also set to zero. After solving the governing equations, the pressure drop was measured at two cross sections that are symmetrically located on each side of the arch and separated by $2L_0$. These scales respect the nondimensionalization in Section II. While we recognize that our theory, which is based on Stokes equations, is strictly valid only when $\mathcal{R}e=0$, we have highlighted that for problems with two distinct length scales ($h_0$ and $L_0$) the ratio of the inertia to the viscous terms in the Navier-Stokes equations involves the product of $\mathcal{R}e$ and $\delta$. Thus, we also performed some numerical calculations with finite $\mathcal{R}e<20$ to show that the results are still useful for finite Reynolds numbers.
For example, simulation results for $\mathcal{R}e=1, \delta=1$, and different $\lambda$ are shown in Fig. \ref{fig:SimResults}. Since the flow rate is constant, the pressure drop increases rapidly as the gap becomes smaller (alternatively, $\lambda$ increases).


\begin{figure}
\vspace{-6mm}
\begin{center}
\resizebox{.67\columnwidth}{!} {\includegraphics{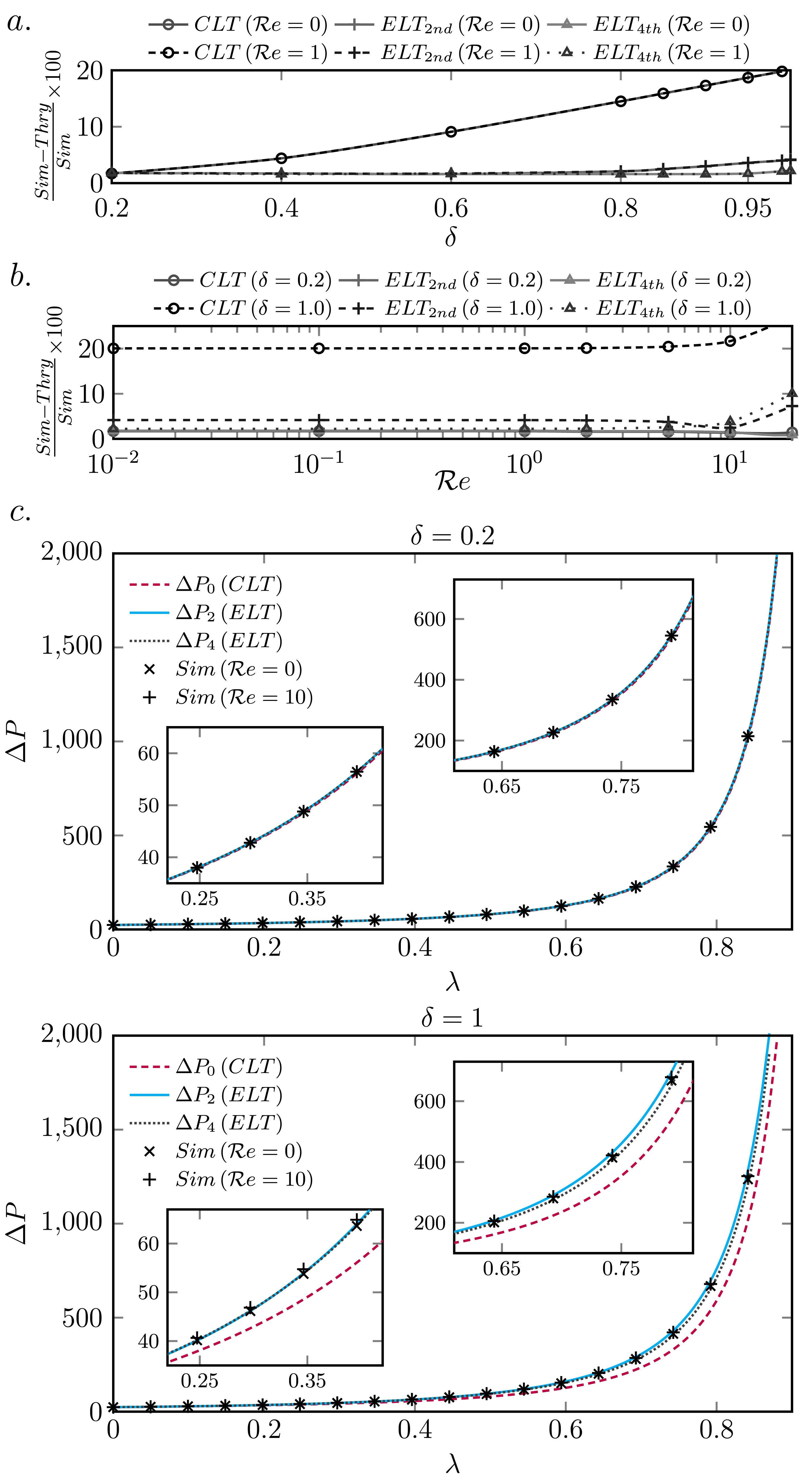}}
\vspace{-8mm}
\end{center}
\caption[]{{\sl a.} and {\sl b.} Deviation of different orders of the analytical solutions from the simulation results while varying $\delta$ and $\mathcal{R}e$, respectively. Each point corresponds to the maximum difference between the simulation and analytical results when changing $\lambda$ from 0 to 0.99. We note that as $\delta \rightarrow 1$, the deviation of CLT from the simulation results becomes significant (about $20\%$) while higher-order analytical solutions are still in a very good agreement with the simulation results for any $\mathcal{R}e \le10$. {\sl c.} Comparison of the dimensionless pressure drop across the arch showing the analytical solutions and simulation results when $\delta = 0.2$ and $\delta=1$.
\label{fig:SimTheory}}
\end{figure}
The comparison between simulation results and different orders of the analytical solutions for a channel with a shape function provided in \eqref{eq:ShapeFunction} is shown in Fig. \ref{fig:SimTheory}. As expected from textbook discussions of classical lubrication theory (see also \citet{Pozrikidis1987}), when $\delta$ is small, $i.e. \,\, \delta \le 0.2$, the CLT as well as higher orders ELT estimate the pressure drop accurately (the errors are within 5\% when $\mathcal{R}e \le10$). By increasing $\delta \rightarrow 1$, the CLT estimations deviate significantly from the simulation results (about 20 \%) while higher orders ELT are still in very good agreement with the simulation results, \textit{i.e.}, three terms in the asymptotic expansion in $\delta$ (including terms $\mathcal{O}\left (\delta^4\right )$) provides very good results even when $\delta \rightarrow 1$ (Fig. \ref{fig:SimTheory}a). We also note that if a maximum $4\%$ error is acceptable, the second-order ELT would be adequate to estimate the pressure drop for a simple shape like \eqref{eq:ShapeFunction} while $\delta \le 1$ and $\mathcal{R}e \le 10$. In the literature on pressure-driven flow in slowly varying channel shapes, it has been shown that domain perturbations have a limited range of utility and so are ``of limited practical interest" \citep[p. 507--508]{Pozrikidis1987}, while here we see that ELT is, in fact, successful even when the wall slopes are order one. As we further increase the Reynolds number (Fig. \ref{fig:SimTheory}b), even the higher orders of the analytical solutions do not estimate the pressure drop accurately as inertial forces become more dominant than the viscous forces and can no longer be ignored. 

\begin{figure}
\begin{center}
\resizebox{0.79\columnwidth}{!} {\includegraphics{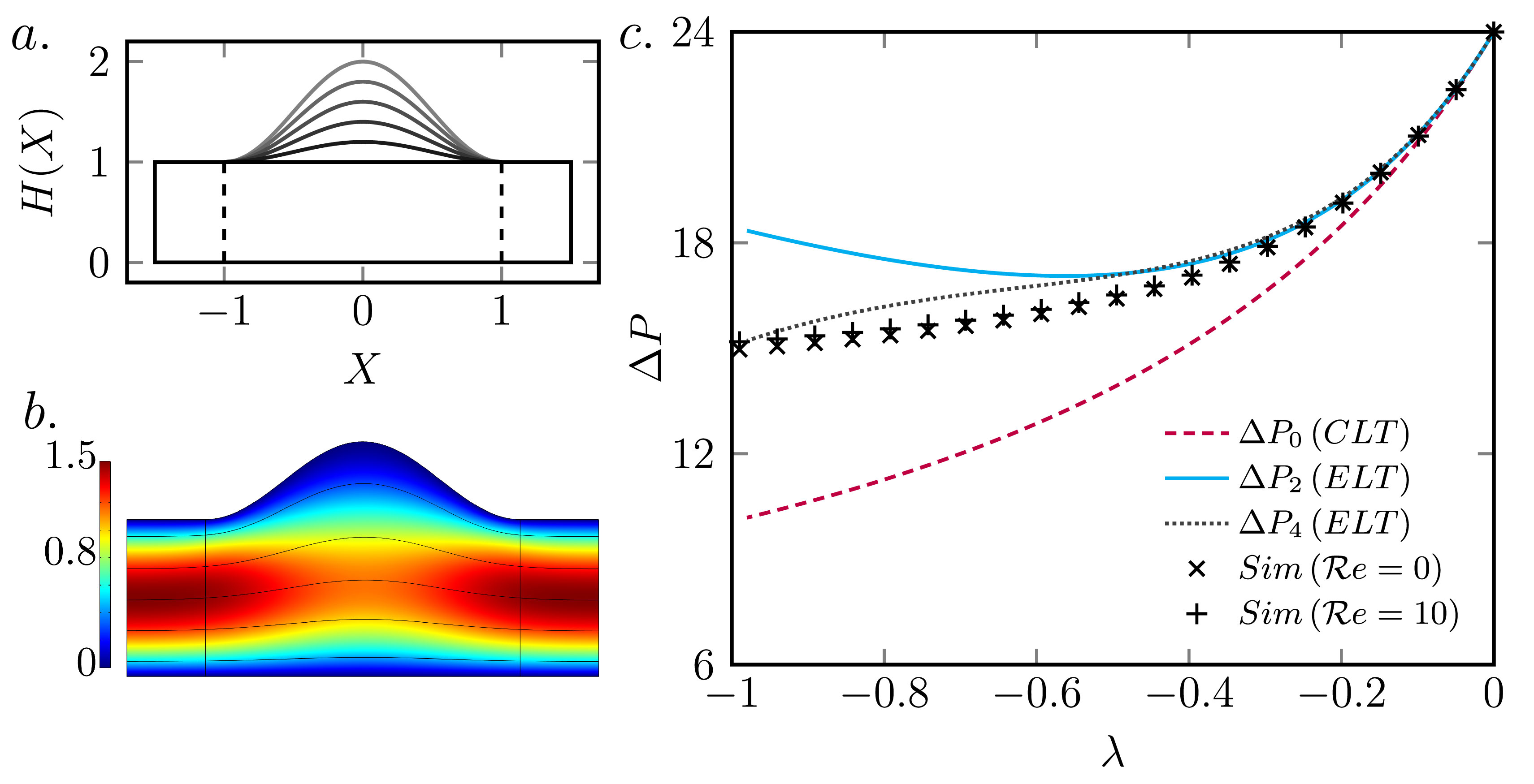}}
\end{center}
\vspace{-8mm}
\caption[]{{\sl a.} A schematic of a channel with a convex shape. The bulge becomes larger as $\lambda \rightarrow -1$. {\sl b.} Velocity magnitude of a channel with 50\% convexity ($\lambda = -\frac{1}{2}$), when $\delta=1$ and $\mathcal{R}e=1$. {\sl c.} Comparison of the dimensionless pressure drop between simulation results and different orders of the analytical solutions for a channel with a convexity shape provided in {\sl a}, when $\delta=1$.
\label{fig:NegShape}}
\end{figure}

We then investigate the use of higher-order ELT in applications where a channel, instead of an obstruction, has a convex shape. We choose the shape function \eqref{eq:ShapeFunction} while varying $\lambda$ from 0 (no bulge) to -1 (bulge with the size of the channel height) (Fig. \ref{fig:NegShape}a). This convexity alters the flow profile (Fig. \ref{fig:NegShape}b) and reduces the pressure drop within the channel. We performed numerical analyses for different $\delta, \mathcal{R}e,\, \text{and}\, \lambda$ and the comparisons are shown in Fig. \ref{fig:NegShape}c. Since we have shown that even for $\delta =1$ the extended lubrication theory provides a reasonable approximation to the full numerical simulations (Fig. \ref{fig:SimTheory}a), here we choose $\delta=1$. For a channel with a sharp convexity, only fourth-order ELT and higher may accurately estimate the pressure drop (within 5\% error) while CLT and the second-order ELT estimations differ from the simulation results by about 30\% and 20\%, respectively. Therefore, higher orders of the analytical solutions significantly improve the estimation of pressure drop within a channel with a significant change in geometry.

\begin{figure}
\begin{center}
\resizebox{0.99\columnwidth}{!} {\includegraphics{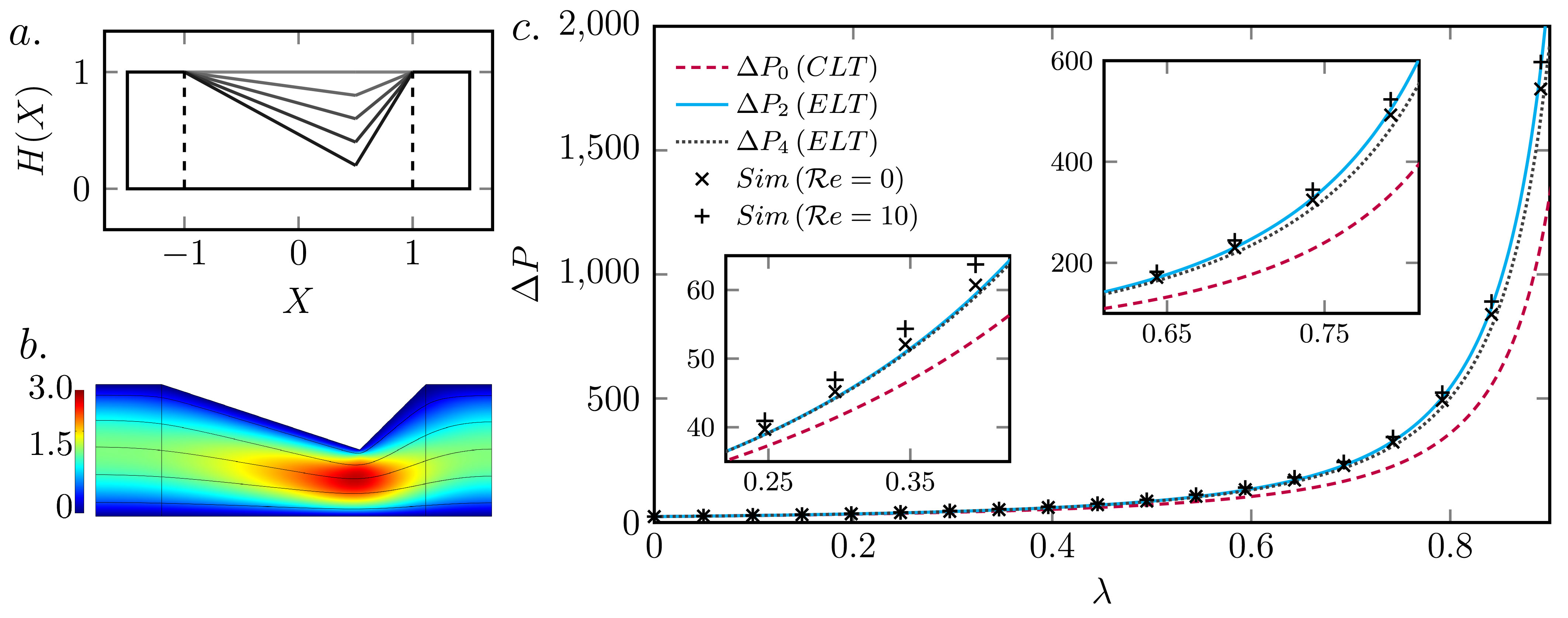}}
\end{center}
\vspace{-6mm}
\caption[]{{\sl a.} A schematic of a channel with a shape function that is not differentiable at $\gamma =0.75$ (see equation \eqref{eq:pw-shapefunc} for the definition of $\gamma$). The gap size decreases as $\lambda \rightarrow 1$. {\sl b.} Velocity magnitude of a channel with a non-differentiable point at $\gamma=0.75$, when $\delta=1$ and $\mathcal{R}e=1$. {\sl c.} Comparison of the dimensionless pressure drop between simulation results and different orders of the analytical solutions for a channel with a single non-differentiable point, provided in {\sl a}, when $\delta=1$.
\label{fig:ContShape}}
\end{figure}

Until now, we have used shape functions that are entirely differentiable. This condition may not be met in models of all applications. Here we provide of an example showing that ELT can be applied to a channel whose shape is continuous but piece-wise differentiable, i.e. the shape function may have finite non-differentiable points. A channel with such a shape function can be divided into smaller sections where each part has a differentiable shape. The pressure drop within each piece is estimated using ELT. Since the flow is laminar, the total pressure drop across the original channel is the summation of the pressure drops within each part. For example, consider a shape function with a single non-differentiable point as follows
\begin{equation}
H(X) = \left\{ \begin{array} {llc}
1-\frac{\lambda}{2\gamma}(X+1)  &\,\, &-1 \leq X \leq 2\gamma-1, \\
1+\frac{\lambda}{2(1-\gamma)}(X-1)  &\,\, &2\gamma-1 < X \leq 1, 
\end{array}\right.
\label{eq:pw-shapefunc}
\end{equation}
where $0<\gamma<1$ is a dimensionless parameter that determines the location of the discontinuity in slope and $1-\lambda$ gives the minimum gap height. This shape function is plotted for $\gamma = 0.75$ in Fig. \ref{fig:ContShape}a. Following the same procedure introduced in Section II, the pressure drop is

\begin{equation}
\Delta P = \Delta P_0 \left(1+\frac{4}{5}\, \lambda^2 \, \delta^2 - \frac{64}{225} \, \lambda^4 \, \delta^4 + \mathcal{O} \left(\delta^6\right) \right)
\label{eq:pw-pressuredrop}
\end{equation}
where $\Delta P_0 = \frac{12\,(2-\lambda)}{(1-\lambda)^2}$, and the terms inside the parentheses correspond to CLT, second-order ELT, fourth-order ELT, and so on, respectively.
We used the same shape function \eqref{eq:pw-shapefunc} to perform numerical simulations, and the comparison is shown in Fig. \ref{fig:ContShape}b. For channels with small $\delta$ ($\delta \le 0.2$), theoretical and numerical results are in good agreement, while for channels with $\delta \approx 1$, pressure drop estimated using the higher orders of the analytical solutions follow the numerical results more closely. These results show that ELT can be applied to a channel with a piece-wise differentiable shape function. We note that shape functions can appear on both sides of a channel and the same procedure can be followed to find analytical solutions at different orders.

\section{Conclusion}
We extended the lubrication approximation by obtaining higher-order terms in a systematic perturbation analysis and compared the analytical results with experiment and numerical simulations. Experimental results were closer to higher-order analytical solutions when the gap was narrow so that the two-dimensional approximation was appropriate. Very good agreement was found between higher-order analytical solutions and the simulation results, confirming that for channels with a high aspect ratio, the higher-order terms of the extended lubrication theory results in a significant improvement in accuracy as compared to the classical lubrication theory. Moreover, though domain perturbations are unsuccessful when the shape changes are modest, extended lubrication theory has been shown to be quite successful, at least for the channel shapes studied here.  Nevertheless, if the wall slope is too steep, we can expect this lubrication approach to fail. Finally, for low Reynolds numbers, simple piece-wise differentiable shape functions can be used with the analytical solutions obtained in this study, which provides a robust tool to accurately estimate the pressure drop in a channel with positive or negative constrictions, whose changes in height are comparable to its length.

\section* {Acknowledgments}

B.T. and D.P.H. are grateful to the National Science Foundation (CMMI--1505125) for financial support. 

\vspace{-1mm}

\section* {References}

\biboptions{numbers,sort&compress}

\end{document}